# AN ALGORITHM FOR IMPROVING THE QUALITY OF COMPACTED JPEG IMAGE BY MINIMIZES THE BLOCKING ARTIFACTS

## Sukhpal Singh[1]

[1]Department of Computer Science and Engineering, Thapar University, Patiala, India.
ssgill@hotmail.co.in

## ABSTRACT

*The Block Transform Coded, JPEG- a lossy image compression format has been used to keep storage and bandwidth requirements of digital image at practical levels. However, JPEG compression schemes may exhibit unwanted image artifacts to appear - such as the 'blocky' artifact found in smooth/monotone areas of an image, caused by the coarse quantization of DCT coefficients. A number of image filtering approaches have been analyzed in literature incorporating value-averaging filters in order to smooth out the discontinuities that appear across DCT block boundaries. Although some of these approaches are able to decrease the severity of these unwanted artifacts to some extent, other approaches have certain limitations that cause excessive blurring to high-contrast edges in the image. The image deblocking algorithm presented in this paper aims to filter the blocked boundaries. This is accomplished by employing smoothening, detection of blocked edges and then filtering the difference between the pixels containing the blocked edge. The deblocking algorithm presented has been successful in reducing blocky artifacts in an image and therefore increases the subjective as well as objective quality of the reconstructed image.*

## 1. INTRODUCTION

As the usage of computers continue to grow, so too does our need for efficient ways for storing large amounts of data (images). For example, someone with a web page or online catalog – that uses dozens or perhaps hundreds of images-will more likely need to use some form of image compression to store those images. This is because the amount of space required for storing unadulterated images can be prohibitively large in terms of cost. Several methods for image compression are available today and these are categorized as: lossless and lossy image compression. The JPEG is a widely used form of lossy image compression standard that centers on the Discrete Cosine Transform (DCT). The DCT works by separating images into parts of differing frequencies. During a step called quantization, where part of compression actually occurs, the less important frequencies are discarded. Then, only the important frequencies remain and are used to retrieve the image in the decompression process. As a result, the reconstructed images contain some distortions. At low bit-rate or quality, the distortion called blocking artifact is unacceptable [1]. This paper work deals with reducing the extent of blocking artifacts in order to enhance the both subjective as well as objective quality of the decompressed image.

## 1.1 Motivation

JPEG defines a "baseline" lossy algorithm, plus optional extensions for progressive and hierarchical coding. Most currently available JPEG hardware and software handles only the baseline mode. It contains a rich set of capabilities that make it suitable for a wide range of applications involving image compression. JPEG requires little buffering and can be efficiently implemented to provide the required processing speed for most cases. Best Known lossless compression methods can compress data about 2:1 on average. Baseline JPEG (color images at 24 bpp) can typically achieve 10:1 to 20:1 compression without visible loss and 30:1 to 50:1 compression visible with small to moderate defects. For Gray Images (at 8 bpp), the threshold for visible loss is often around 5:1 compression. The baseline JPEG coder is preferable over other standards because of its low complexity, efficient utilization of memory and reasonable coding efficiency. Although more efficient compression schemes do exist, but JPEG is being used for a long period of time that it has spread its artifacts over all the digital images [6]. The need for a blocking artifact removal technique is therefore a motive that constantly drives new ideas and implementations in this field. Considering the wide spread acceptance of JPEG(baseline) standard [16], this paper suggested a post processing algorithm that does not make any amendments into the existing standard, and reduces the extent of blocking artifacts. That is, the work is being done to improve the quality of the image.

## 1.2 Paper Outline

This paper deals with three of the image processing operations:

1.2.1 **Image restoration-** Restoration [12] takes a corrupted image and attempts to recreate a clean original. It is clearly explained in the figure 1.1.

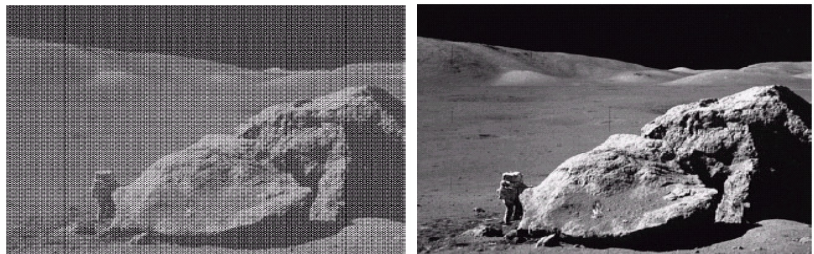

Figure 1.1a) Original image          b) Image after restoration

1.2.2 **Image Enhancement -** Image Enhancement alters an image to makes its meaning clearer to human observers [13]. It is often used to increase the contrast in images that are overly dark or light explained in the figure 1.2.

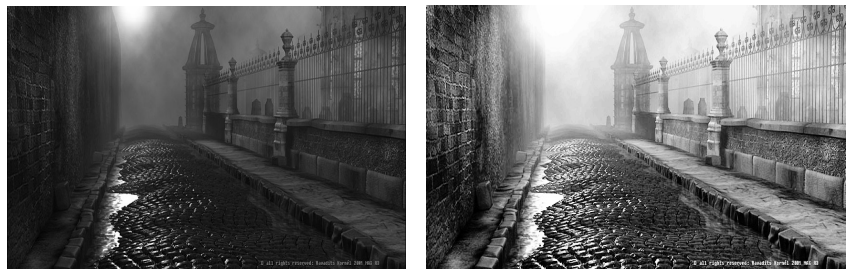

Figure 1.2 a) Original image          b) Image after enhancement

**1.2.3 Image Compression-** Image compression is the process that helps to represent image data with as few bits as possible through exploiting redundancies in the data while maintaining an appropriate level of quality for the user [14]. Lossy compression is shown in figure-1.3.

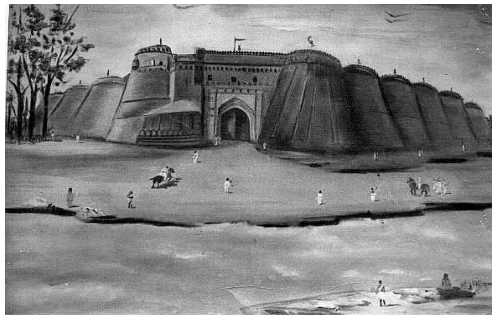 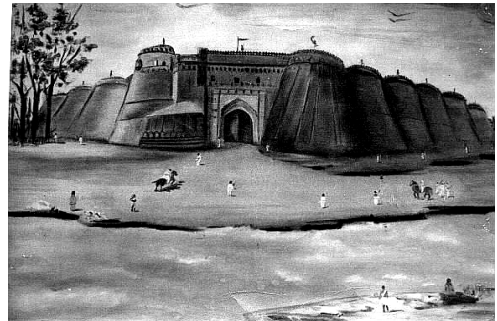

Figure 1.3 a) Original image        b) Image after compression

## 1.3 Image Compression

As the beginning of the third millennium approaches, the status of the human civilization is best characterized by the term "Information Age" [5]. Information, the substance of this new world, despite its physical non-existence can dramatically change human lives. In a sense, being in the right place at the right time depends on having the right information, rather than just having luck. Information is often stored and transmitted as digital data [6]. However, the same information can be described by different datasets in figure 1.4.

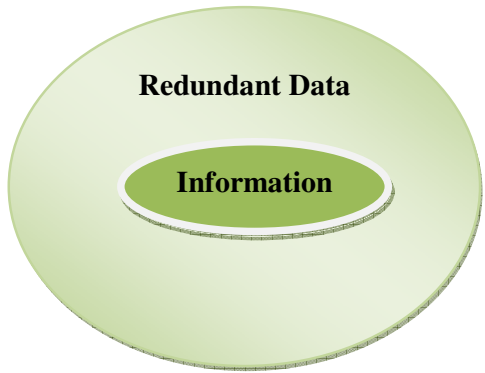

DATA = REDUNDANT DATA + INFORMATION

Figure-1.4: Relationship between data and information [6]

The shorter the data description, usually the better, since people are interested in the information and not in the data. Compression is the process of transforming the data description into a more succinct and condensed form. Thus improves the storage efficiency, communication speed, and security [9]. Compressing an image is significantly different than compressing raw binary data. This is because images have certain statistical properties which can be exploited by encoders specifically designed for them.

## 1.4 Motivation behind image compression [11]

A common characteristic of most images is that the neighboring pixels are correlated and therefore contain redundant information [20]. The foremost task then is to find less correlated representation of the image. In general, three types of redundancy can be identified:

**1.4.1 Coding Redundancy:** If the gray levels of an image are coded in a way that uses more code symbols than absolutely necessary to represent each gray level, the resulting image is said to have coding redundancy.

**1.4.2 Interpixel Redundancy:** This redundancy is directly related to the interpixel correlations within an image. Because the value of any given pixel can be reasonably predicted from the value of its neighbors, the information carried by individual pixels is relatively small. Much of the visual contribution of a single pixel to an image is reduntant; it could have been guessed on the basis of the values of its neighbors.

**1.4.3 Psychvisual Redundancy:** This redundancy is fundamentally different from other redundancies. It is associated with real or quantifiable visual information [21]. Its elimination is possible only because the information itself is not essential for normal visual processing. Since the elimination of psychvisually redundant data results in a loss of quantitative information, it is commonly referred to as quantization.

### 1.5 Image compression model

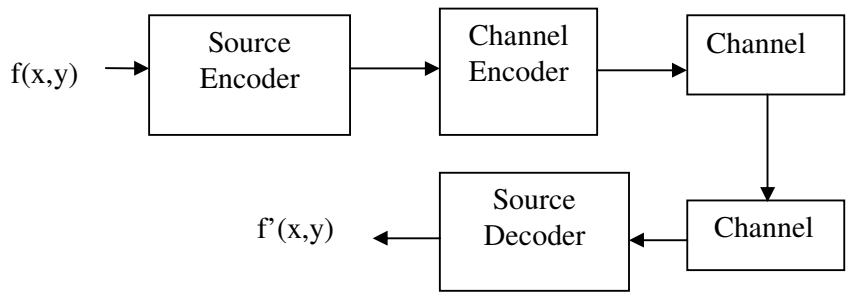

Figure-1.5: Image compression Model [16]

As the above figure shows, a compression system consists of two distinct structural blocks: an encoder and a decoder. An input image f(x,y) is fed into the encoder, which creates a set of symbols from the input data. After transmission over the channel, the encoded representation is fed to the decoder, where the reconstructed output image f'(x,y) is generated. In general, f'(x,y) may or may not be the exact replica of f(x,y).The encoder is made up of a source encoder, which removes input redundancies, and a channel encoder, which increases the noise immunity of the source encoder's output. Same is in the case of decoder, but functions in reverse direction explained in figure 1.5.

### 1.5 Compression Techniques

There are two different ways to compress images-lossless and lossy compression.

#### 1.6.1 Lossless Image Compression

A lossless technique means that the restored data file is identical to the original explained in figure 1.6. This type of compression technique is used where the loss of information is unacceptable [22]. Here, subjective as well as objective qualities are given importance. In a nutshell, decompressed image is exactly same as the original image.

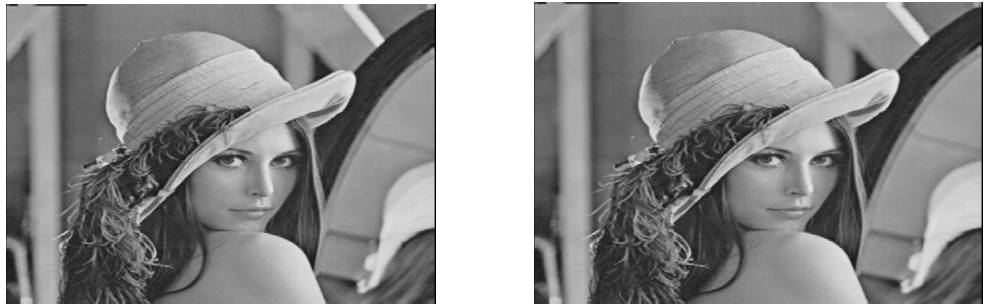

a). Original Image                    b).Decompressed Image
Figure-1.6: Relationship between input and output of Lossless Compression

### 1.6.2 Lossy Image Compression

It is based on the concept that all real world measurements inherently contain a certain amount of noise. If the changes made to these images, resemble a small amount of additional noise, no harm is done [23]. Compression techniques that allow this type of degradation are called lossy explained in figure 1.7. The higher the compression ratio, the more noise added to the data. In a nutshell, decompressed image is as close to the original as we wish.

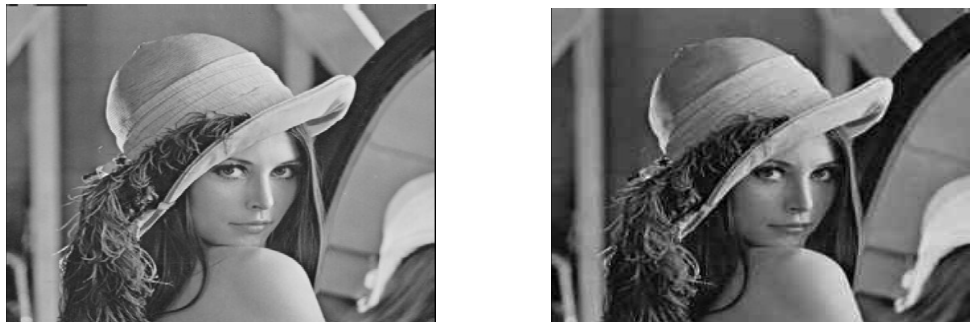

a).Original Image                    b).Decompressed Image
Figure-1.7: Relationship between input and output of Lossy Compression

Lossless compression technique is reversible in nature, whereas lossy technique is irreversible. This is due to the fact that the encoder of lossy compression consists of quantization block in its encoding procedure.

## 1.6 JPEG

JPEG (pronounced "jay-peg") is a standardized image compression mechanism. JPEG also stands for Joint Photographic Experts Group, the original name of the committee that wrote the standard. JPEG is designed for compressing full-color or gray-scale images of natural, real-world scenes. It works well on photographs, naturalistic artwork, and similar material [18]. There are lossless image compression algorithms, but JPEG achieves much greater compression than with other lossless methods. JPEG involves lossy compression through *quantization* that reduces the number of bits per sample or entirely discards some of the samples. The usage of JPEG compression method is motivated because of following reasons:-

**1.7.1** The *compression ratio* of lossless methods is not high enough for image and video compression.

**1.7.2** JPEG uses *transform coding*, it is largely based on the following observations:

*Observation 1:* A large majority of useful image contents change relatively slowly across images, i.e., it is unusual for intensity values to alter up and down several times in a small area, for example, within an 8 x 8 image block [24].

*Observation 2:* Generally, lower spatial frequency components contain more information than the high frequency components which often correspond to less useful details and noises [26].

Thus, JPEG is designed to exploit known limitations of the human eye, notably the fact that small color changes are perceived less accurately than small changes in brightness. JPEG can vary the degree of lossiness by adjusting compression parameters [22]. Useful JPEG compression ratios are typically in the range of about 10:1 to 20:1. Because of the mentioned plus points, JPEG has become the practical standard for storing realistic still images using lossy compression.JPEG (encoding) works as shown in the figure 1.8. The decoder works in the reverse direction. As quantization block is irreversible in nature, therefore it is not included in the decoding phase. It is clearly explained in the figure 1.8.

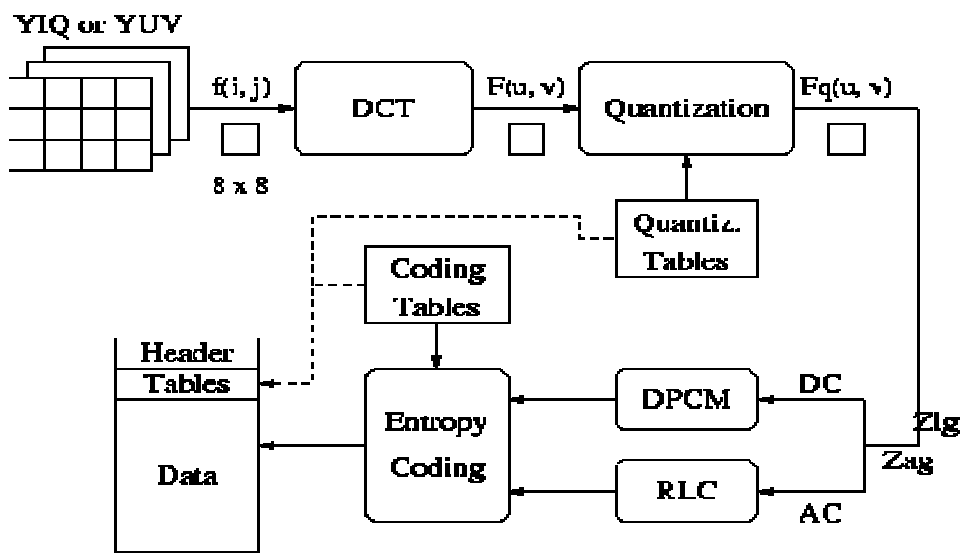

Figure 1.8: Steps in JPEG Compression [22]

A major drawback of JPEG (DCT-based) is that blocky artifacts appear at low bit-rates in the decompressed images. Such artifacts are demonstrated as artificial discontinuities between adjacent image blocks. An image illustrating such blocky artifacts is shown in the figure below:-

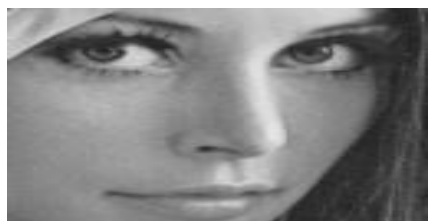 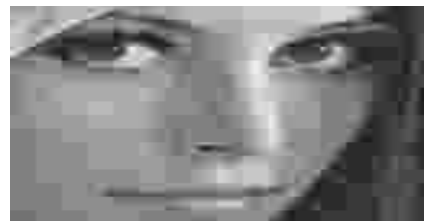

Figure1.9: a).Actual Image      b).Blocked Image

## 2. LITERATURE SURVEY

In practice, it is always a tradeoff between the coding bit rate and the coded image quality. Generally speaking, increasing coding bit rate can improve the quality of the reconstructed image, but it is limited by channel bandwidth or storage capacity. In order to achieve specified bit rate, a large quantization step for transformed coefficients should be adopted, so the blocking artifacts are more obvious at high compressed ratios, which often limit the maximum compression capacity that can achieve (Wang, 2004). Many algorithms have been proposed for reducing these blocking artifacts. These algorithms can be classified into two major categories. One is to use different encoding schemes, such as the interleaved block transform, the lapped transform, and the combined transform. Further, post-processing algorithms can be categorized as spatial, frequency (DCT-based) and combined (hybrid) algorithms based on the domain in which they operate. The research done by various authors in the same field have been summarized in Table 2.1.

Table 2.1: Summary of research done by different authors

| Sr. | Author's name | Paper Title | Focus |
|---|---|---|---|
| 1. | Ramamurthi and Gersho | space-variant filter that adapts to local characteristics of the signal | The algorithm distinguishes edge pixels from non-edge pixels via a neighborhood testing and then switches between a 1D filter and a 2D filter accordingly to reduce blocking effects [11]. |
| 2. | Long | Adaptive Deblocking of Images with DCT Compression | Proposed a very unique concept of "noise injection." By adding a small amount of random noise to a blocky image in the DCT domain (the "noise injection" process), he converted the blocky image into an image more like being corrupted with "Gaussian-type" noise, thus changed the image deblocking problem into a denoising problem [3, 24]. |
| 3. | Robertson | DCT Quantization Noise in Compressed Images | Provides a spatial domain model of the quantization error based on statistical noise model of the error introduced when quantizing the DCT coefficients [4, 22]. |
| 4. | Tuan | Blocking artifacts removal by a hybrid filter method | Suggested a method that simultaneously performs an edge-preserving and a low-pass filtering of the degraded image [19]. |
| 5. | Aria | Enhancement of JPEG-Compressed Images by Re-application of JPEG | Simply re-applies JPEG to the shifted versions of the already-compressed image, and forms an average [2]. |
| 6. | Kwan | Blocking Artifacts Reduction Algorithm in Block Boundary area using Neural Network | Proposed an algorithm using block classification and feed forward neural network filters in the spatial domain [12]. |
| 7. | Park | Blocking Artifacts Reduction in Block-Coded Images Using Self-Similarity | Proposed algorithm considering piecewise self-similarity within different parts of the image as a priori to give reasonable modification to the block boundary pixels [17]. |

| 8. | Gothundarumun | Total Variation for the Removal of Blocking Effects in DCT Based Encoding | Proposed a non-linear method that reduces artifacts by minimizing the total variation via a level set formulation [7]. |
|---|---|---|---|
| 9. | Wenfeng | A De-Blocking Algorithm and a Blockiness Metric for Highly Compressed Images | Proposed a de-blocking algorithm based on the number of connected blocks in a relatively homogeneous region, the magnitude of abrupt changes between neighboring blocks, and the quantization step size of DCT coefficients [3, 26]. |
| 10. | Averbuch | Deblocking of Block-Transform Compressed Images Using Weighted Sums of Symmetrically Aligned Pixels | Appllied weighted sums on pixel quartets, which are symmetrically aligned with respect to block boundaries [3]. |
| 11. | Yang | Regularized Reconstruction to Reduce Blocking Artifacts of Block Discrete Cosine Transform Compressed Images | Suggested two methods for solving this regularized problem. The first was based on the theory of projections onto convex sets (POCS) while the second was based on the constrained least squares (CLS) approach [28]. For the POCS-based method, a new constraint set was defined that conveys smoothness information not captured by the transmitted B-DCT coefficients [27], and the projection onto it was computed. For the CLS method an objective function was proposed that captures the smoothness properties of the original image [4]. |
| 12. | Triantafyllidis | Blockiness Reduction in JPEG Coded Images | Suggested an algorithm that firstly estimates the AC coefficients based on their observed probability distribution and then, a postprocessing scheme consisting of a region classification algorithm and a spatial adaptive filtering is applied for blockiness removal [9, 22]. |
| 13. | Hyuk | Blocking-Artifact Reduction in Block-Coded Images Using Wavelet-Based Subband Decomposition | Proposed a post-processing method in the wavelet transform domain. Knowing that sub band coding does not suffer from blocky noise, the proposed technique is designed to work in the sub band domain [8]. |
| 14. | Triantafyllidis | Detection of Blocking Artifacts of Compressed Still Images | Detects the regions of visible blocking artifacts and uses the estimated relative quantization error calculated when the DCT coefficients are modeled by a Laplacian probability function [18]. |
| 15. | Zhao | Postprocessing technique for blocking artifacts reduction in DCT domain | Stated that the DCT distributions of differently shifted blocks before quantization are approximately identical, whereas the DCT distributions of differently shifted blocks in the compressed image are considerably different and used the difference to expose the blocking artifacts [18]. |
| 16. | Wang | Adaptive Reduction of Blocking Artifacts in DCT Domain for | Removes these discontinuities by Walsh transform and local threshold technology, the precision of edge detection by Sobel operator is |
| | | Highly Compressed Images | improved and blurring in objective edge is reduced [20-21]. |
| 17. | Alan's | Blocking Artifacts Suppression in Block-Coded Images Using Overcomplete Wavelet Representation | Suggested a non-iterative, wavelet-based deblocking algorithm exploiting the fact that block discontinuities are constrained by the dc quantization interval of the quantization table [4], as well as the behavior of wavelet modulus maxima evolution across wavelet scales to derive appropriate threshold maps at different wavelet scales [13]. |
| 18. | Kizewski's | Image Deblocking Using Local Segmentation | Suggested an image deblocking filter by employing local segmentation techniques prior to applying a simple value averaging model to a local neighborhood of 3x3 DCT coefficients [11]. |
| 19. | Wang | Fast Edge-Preserved Postprocessing for Compressed Images | Proposed a fast algorithm based on the concept to decompose a row or column image vector to a gradually changed signal and a fast variation signal [25]. |
| 20. | Nallaperumal | Removal of blocking artifacts in JPEG compressed images using Dual Tree Complex Wavelet Filters for Multimedia on the Web | Proposed an algorithm based on the fact that the high frequency [15] details of the coded image are mainly contaminated by quantization noise [10]. |
| 21. | Luo | Removing the Blocking Artifacts of Block-Based DCT Compressed Images | Proposed an adaptive approach. For smooth regions, the method takes advantage of the fact that the original pixel levels in the same block provide continuity and use this property and the correlation between the neighboring blocks to reduce the discontinuity of the pixels across the boundaries [14]. |

# 3. PRESENT WORK
## 3.1 Problem Formulation

In JPEG (DCT based) compresses image data by representing the original image with a small number of transform coefficients. It exploits the fact that for typical images a large amount of signal energy is concentrated in a small number of coefficients. The goal of DCT transform coding is to minimize the number of retained transform coefficients while keeping distortion at an acceptable level.In JPEG; it is done in 8X8 non overlapping blocks. It divides an image into blocks of equal size and processes each block independently. Block processing allows the coder to adapt to the local image statistics, exploit the correlation present among neighboring image pixels, and to reduce computational and storage requirements. One of the most degradation of the block transform coding is the "blocking artifact". These artifacts appear as a regular pattern of visible block boundaries. This degradation is a direct result of the coarse quantization of the coefficients and the independent processing of the blocks which does not take into account the existing correlations among adjacent block pixels. In this paper attempt is being made to reduce the blocking artifact introduced by the Block DCT Transform in JPEG.

## 3.2. Objective

The primary objective of this paper is to develop a post-processing algorithm for the reduction of blocking artifacts, as it does not need to modify the existing standard. The algorithm tried to fulfill the following criteria's too.

**3.2.1.** Reduce the extent of blocking artifacts
**3.2.2**. Make efficient utilization of resources
**3.2.3.** Preserves the edges.
**3.2.4**. Do not result in other type of artifacts

## 3.3. Design and Implementation
### 3.3.1 Algorithm

This algorithm is implemented in MatLab 7.There are mainly three segments:

*a) Segment-1*

From the earlier discussion, it has been cleared that in JPEG the image data is dealt block by block basis. That is DCT is applied on each block independently and the same for the quantization .Because of this fact, blocking artifacts came into existence representing the lossy nature of the JPEG standard. This fact is exploited in this segment. As difference exists between each block, an attempt is made (spatially) to smooth out this difference.

There are many smoothing filters that can easily smoothes the difference, but it results in blurring (which is another artifact).The smoothening is being done in a very tactful manner. Due to low quality, edges are quite visible in the image (blocked). Smoothening is done by gradually decreasing the difference between the block edges. The point is also considered that the blocking artifact should not move inside the block, that's why gradual attempt (that is neighboring pixels are also manipulated) is made to reduce the difference. The key idea behind smoothing the blocked image is to reduce the extent of blockiness without blurring the image. Smoothening is done by considering six neighboring pixels on either side of pixels containing the block edge. The algorithm used in this segment is described below:

*Input*: Preprocessed Image

Outp*ut*: Uniformly deblocked Image

*Assumptions:* a(f(x,y)) and b(f(x,y+1)) are two adjacent pixels having block boundary in between, .in a vertical direction.

> *Algorithm:*
> If |a-b|<threshold
//no change in the pixels values.
Else
> > s = |a-b|/2
> > > if a<b
> > > f(x,y)=f(x,y)+s,f(x,y+1)-s
> > > > f(x,y-1)=f(x,y-1)+s/2, f(x,y+2)=f(x,y+2)-s/2
> > > > f(x,y-2)=f(x,y-2)+s/2, f(x,y+3)=f(x,y+3)-s/4
> > > elseif a>b
> > > > f(x,y)=f(x,y)-s,f(x,y+1)+s

26

f(x,y-1)=f(x,y-1)-s/2, f(x,y+2)=f(x,y+2)+s/2
f(x,y-2)=f(x,y-2)-s/2, f(x,y+3)=f(x,y+3)+s/4
else
//nothing
end

The algorithm is repeated for the all the eight pixels along the vertical edge and similar algorithm is used for the horizontal edges.

*b)  Segment-2*

As mentioned, an attempt is made in the previous segment to reduce the blocking artifacts and the artifacts are reduced to some extent. But in the previous segment, blocking artifacts are dealt in an uniform manner. In this segment, the blocked edges are first detected and then deblocked by using Gaussian formula. Here detection and deblocking are interlinked. That is, the pixel being deblocked depends upon the detection criteria.

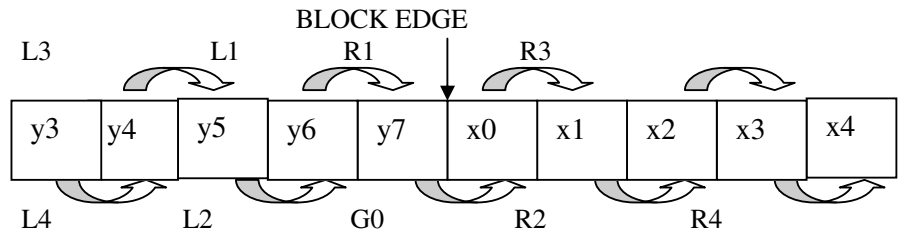

Figure-3.1: Detection Sequence

*Input*: Eight rows of ten pixels having the edge in the centre, which is being detected against blockiness.

Outp*ut*: Deblocked Edge

*Algorithm:*
1.  Initialize counter = 0.
2.  Considering first row.
3.  Assign G0= |x0-y7|     //difference between two boundary pixels.
4.  Assign the difference between each pair of adjacent pixels on left and right-hand side of the block boundary are also calculated and denoted by Li and Ri (i=1,2,3,4) respectively.

   1.  If MAX(L1,L2,L3,L4)<G0 (1) or MAX(R1,R2,R3,R4)<G0   (2)
       // boundary gap is detected
       //current row is marked
       counter= counter+1    // increment the counter
   2.  Repeat the steps 3-5 for the rest seven rows.
   3.  If counter> TH, then blocking artifact is claimed     // TH→ threshold value
   4.  1-D filter using Gaussian formula is applied to {x0,y7.y6}(if equation 1 holds)
       or
       {y7,x0,x1}  (if equation 2 holds)  having  window size=5, along the marked rows.

Detection is done by following the above mentioned detection algorithm for each internal edge of the image and if these follow the decided criteria, only then the edge is marked as BLOCKED edge. The blocked edge is deblocked by applying gaussain formula:

$$Y = \sum_{j=1}^{N} x_j w_j \; / \; ( \; \sum_{j=1}^{N} w_j \; )$$

where $w_j = \exp[-(x_c - x_j)^2 / 2\varepsilon^2]$
N is the window size
$X_C$ is the centre pixel
$\varepsilon$ = average difference from centre pixel

*c)  Segment-3*

Although there is no consensus on a method to measure block artifacts in images, one measure has been used in most of the papers encountered - Peak Signal to Noise Ratio (PSNR). PSNR is basically a logarithmic scale of the mean squared difference between two sets of values (pixel values, in this case). It is used as a general measure of image quality, but it does not specifically measure blocking artifacts. In observed literature, PSNR is used as a source-dependant artifact measure, requiring the original, uncompressed image to compare with. PSNR is defined as:

$PSNR = 20\log_{10}(255/\sqrt{MSE})$
Where $MSE = \sum(B_i - A_i)/n$
where i = 0 to n and n is the number of pixels in the image.

It is easily seen that this blockiness measure is not actually aware of the artifacts it is measuring - it is simply a gauge of how different the corresponding (that is, the same position) pixel values are between two images. Because a blocky image is different from the original and a severely blocky image is more, so PSNR is an acceptable measure, and hence the primary measure used to compare the proposed method. However, two images with completely different levels of perceived blockiness may have almost identical PSNR values. The working of the algorithm is explained by the flowchart drawn in figure 3.2.

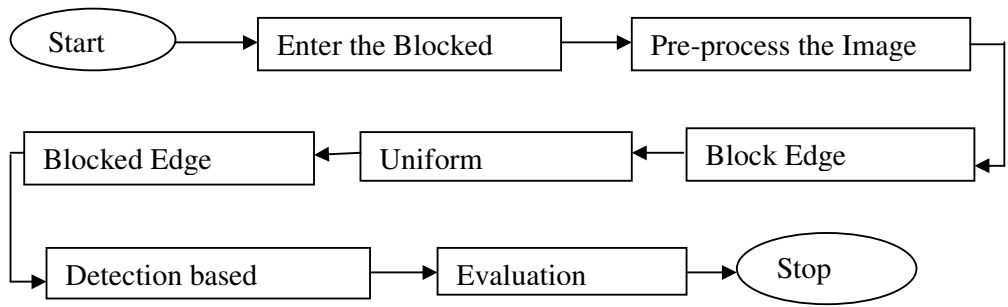

Figure: 3.2. Flowchart of the Proposed Algorithm

## 3.4 Test Images

Seven images have been selected for checking the validation of the proposed algorithm. All are 512X512 .jpg images at different quality factors. It is clearly explained in the figure 3.3 and figure 3.4.

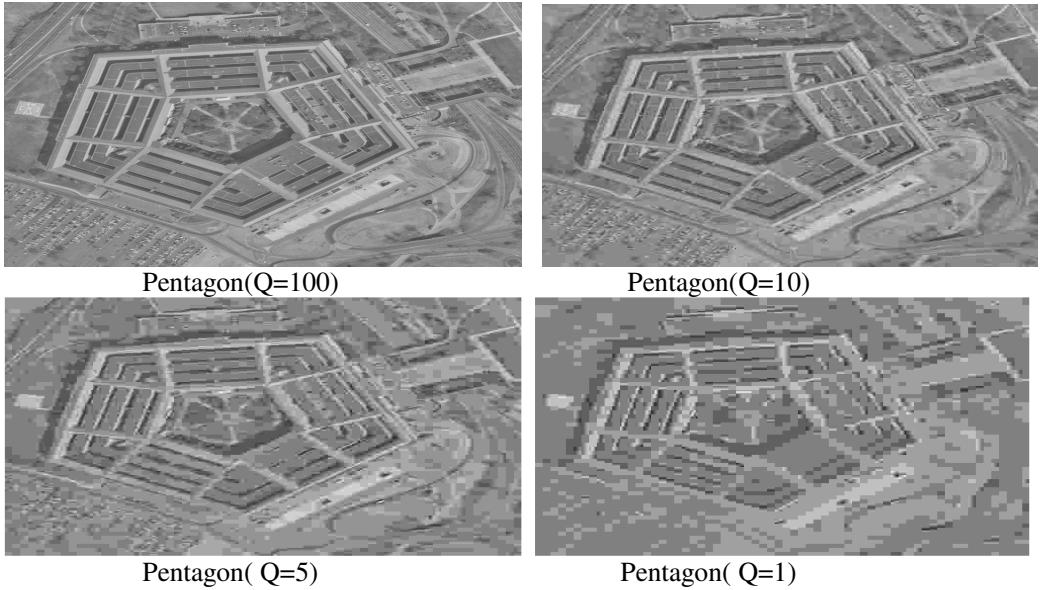

Pentagon(Q=100)        Pentagon(Q=10)

Pentagon( Q=5)        Pentagon( Q=1)

Figure-3.3: Pentagon Image (512X512)

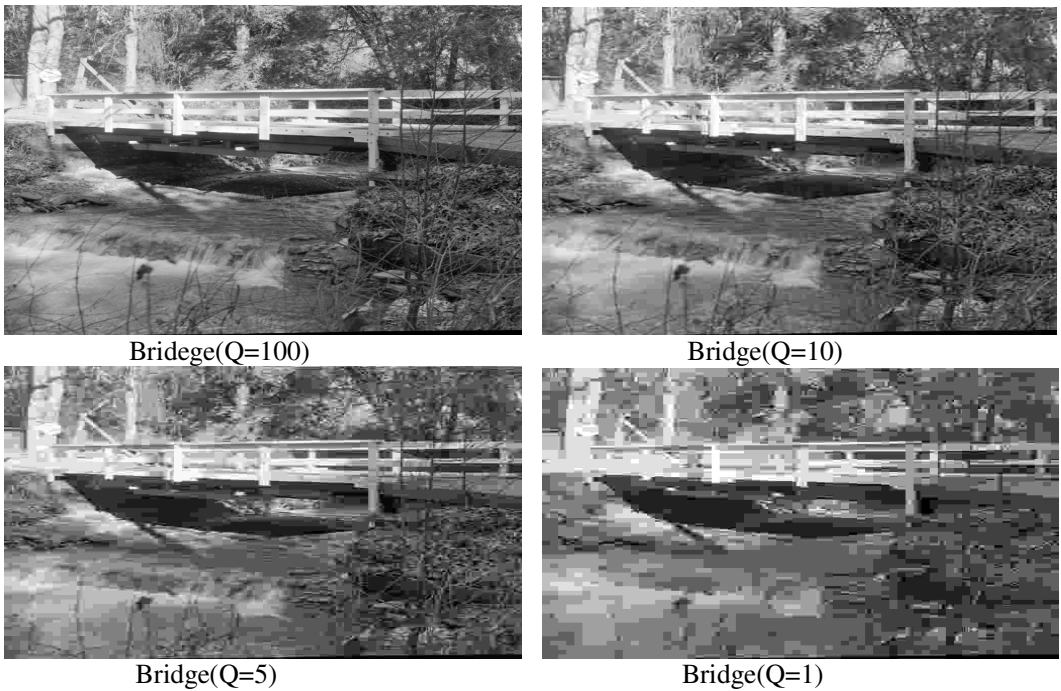

Bridege(Q=100)        Bridge(Q=10)

Bridge(Q=5)        Bridge(Q=1)

Figure-3.4: Bridge Image (512X512)

## 4. RESULTS AND DISCUSSIONS

The proposed algorithm is tested on various images with different characteristics. The algorithm is applied on seven images (as mentioned in the previous chapter) at different quality parameters (Q=10, 5, and 1).The results are shown in Fig: 4.1, 4.2, 4.3 and Table-1.

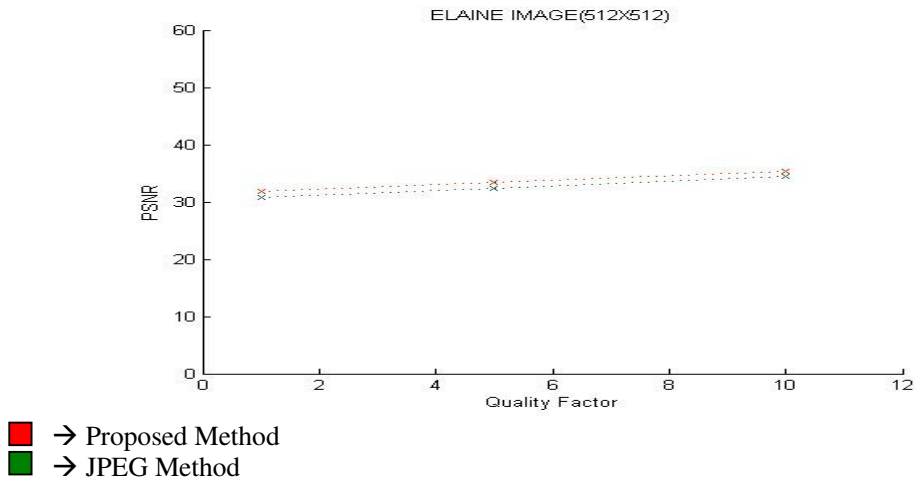

→ Proposed Method
→ JPEG Method

Figure: 4.1 Relationships between PSNR and Quality of Elaine Image

The above figure shows the relationship between PSNR and Quality of Elaine Image using JPEG Method (green) and Proposed Method (red).It is very clear from the plot that there is increase in PSNR value of image with the use of proposed method over the JPEG method. This increase represents improvement in the objective quality of the image.

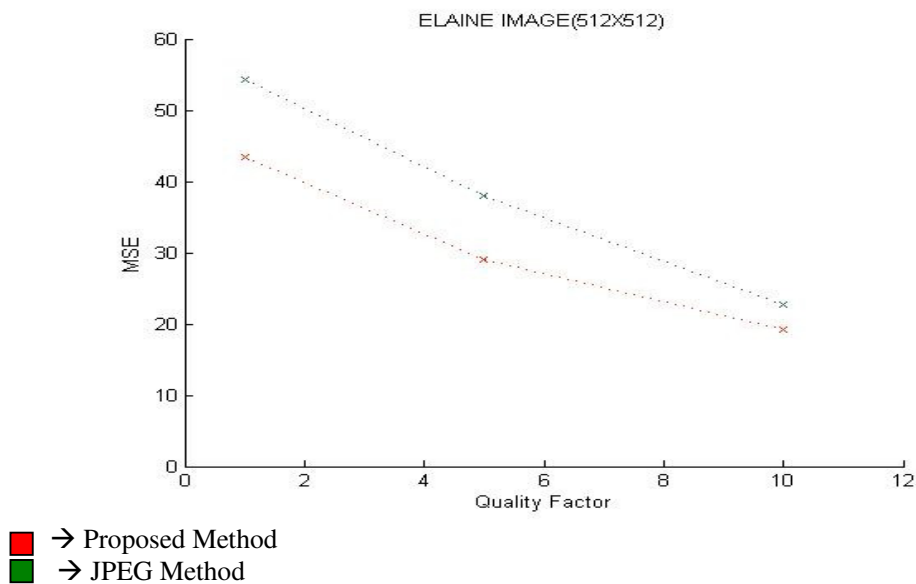

→ Proposed Method
→ JPEG Method

Figure: 4.2 Relationships between MSE and Quality of Elaine Image

The above figure shows the relationship between MSE and Quality of Elaine Image using JPEG Method (green) and Proposed Method (red).It is very clear from the plot that there is decrease in MSE value of image with the use of proposed method over the JPEG method. This decrease represents improvement in the objective quality of the image.

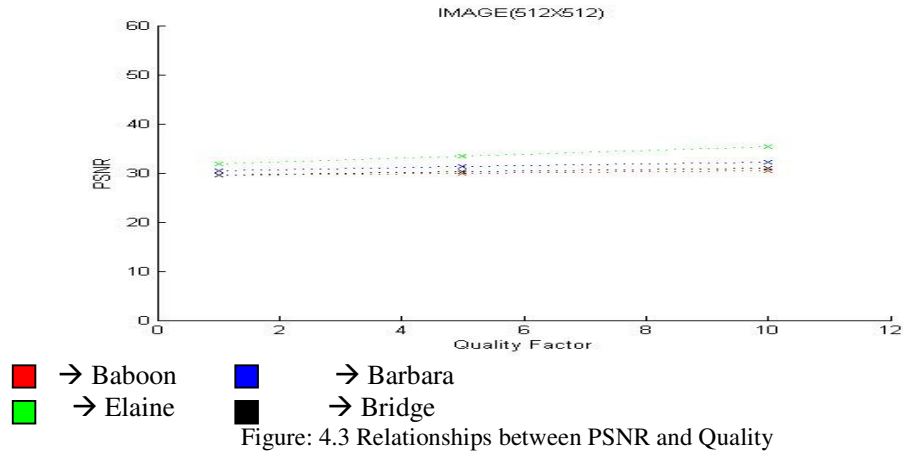

Figure: 4.3 Relationships between PSNR and Quality

Figure 4.3 shows the relationship between PSNR and quality factor of various images using the proposed method. The top-most line (green) representing Elaine image shows that at quality=1, PSNR= 31.7557; at quality=5, PSNR= 33.4897 and at quality=10, PSNR=35.2924.This shows that PSNR value increases with increase in quality of image. This reveals that the relationship between quality and PSNR value of Lena image obtained by the proposed method is in consent with the fact that PSNR value increases with increase in the quality of image. The same fact fulfilled by other images represents the validity of the proposed algorithm. The application of the proposed algorithm on Lena Image (512X512) at different bpp is shown in figure: 4.4-4.6.

**At Quality Parameter = 10(0.25 bpp)**

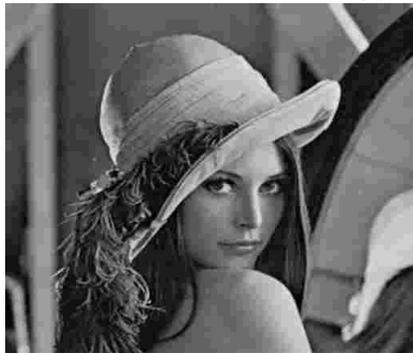
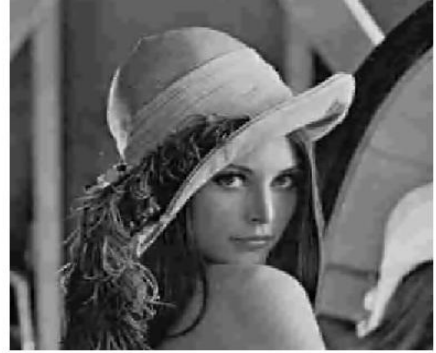

Figure-4.4: a). Blocked Image          Figure-4.4: b) Deblocked Image

**At Quality Parameter = 5(0.18 bpp)**

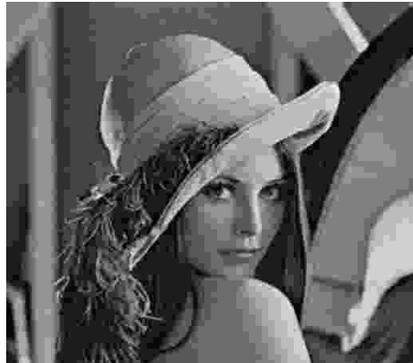
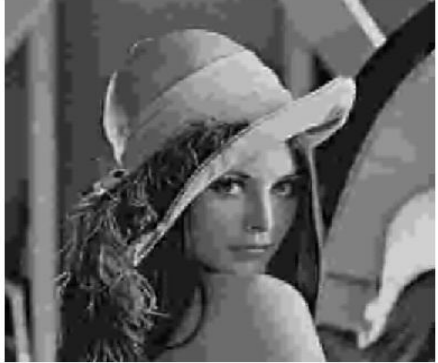

Figure-4.5: a). Blocked Image          Figure-4.5: b). Deblocked Image

**At Quality Parameter = 1(0.1 bpp)**

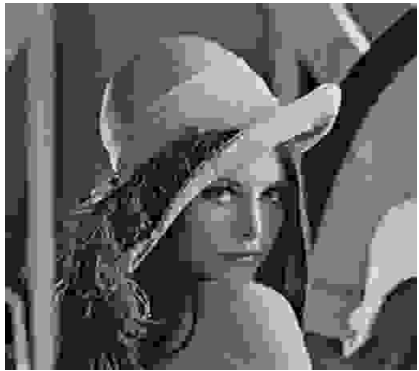 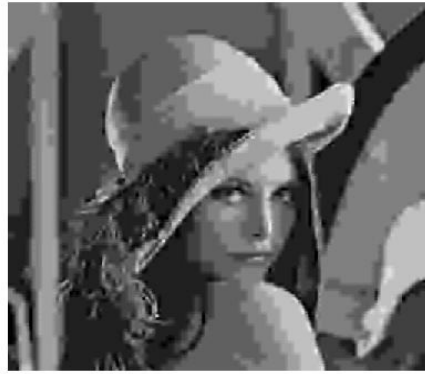

Figure-4.6: a). Blocked Image          Figure- 4.6: b). Deblocked Image

The figure 4.4-4.6 shows the improvement in deblocked image after application of proposed algorithm to the blocked image at various quality (bpp) values. This represents that the proposed algorithm increases the subjective quality of the image. The table below shows the PSNR and MSE values for test images using JPEG and Proposed Method.

Table-1a): Showing improved PSNR and MSE

| Image | Quality | JPEG | | Proposed Algorithm | |
|-------|---------|------|------|------|------|
| | | PSNR | MSE | PSNR | MSE |
| **Barbara** | Q=10 | 32.1245 | 39.8685 | 32.1252 | 39.8624 |
| | Q=5 | 30.8581 | 53.3670 | 31.3020 | 48.1817 |
| | Q=1 | 29.8702 | 66.9973 | 30.3731 | 59.6722 |
| **Peppers** | Q=10 | 34.6592 | 22.2414 | 35.1713 | 19.7675 |
| | Q=5 | 32.1886 | 39.2840 | 33.1044 | 31.8158 |
| | Q=1 | 30.8311 | 53.6992 | 31.8167 | 42.7967 |
| **Baboon** | Q=10 | 30.3658 | 59.7729 | 30.3536 | 59.9406 |
| | Q=5 | 29.6766 | 70.0523 | 29.9106 | 66.3775 |
| | Q=1 | 29.3194 | 76.0573 | 29.5498 | 72.1272 |
| **Elaine** | Q=10 | 34.5543 | 22.7849 | 35.2924 | 19.2239 |
| | Q=5 | 32.3426 | 37.9154 | 33.4897 | 29.1146 |
| | Q=1 | 30.7786 | 54.3522 | 31.7557 | 43.4025 |

Table-1b): Showing improved PSNR and MSE

| Image | Quality | JPEG | | Proposed Algorithm | |
|-------|---------|------|------|------|------|
| | | PSNR | MSE | PSNR | MSE |
| **Bridge** | Q=10 | 30.9221 | 52.5855 | 30.9410 | 52.3582 |
| | Q=5 | 30-0113 | 64.8560 | 30.2752 | 61.0319 |
| | Q=1 | 29.2963 | 74.4631 | 29.5761 | 71.6927 |
| **Pentagon** | Q=10 | 31.9616 | 41.3920 | 32.2033 | 39.1513 |

| | Q=5 | 30.8437 | 53.5437 | 31.3752 | 47.3764 |
|---|---|---|---|---|---|
| | Q=1 | 29.7467 | 68.9296 | 30.1600 | 62.6725 |
| **Lena** | Q=10 | 35.4089 | 18.7149 | 35.7721 | 17.2135 |
| | Q=5 | 32.6314 | 35.4763 | 33.5672 | 28.5993 |
| | Q=1 | 30.9227 | 52.5793 | 31.7384 | 43.5749 |

It is clear from the above table that the algorithm increases the PSNR and decreases the MSE value for quality parameter (Q =10, 5, 1).

# 5. CONCLUSION AND FUTURE SCOPE

It is very much clear from the displayed results that the artifacts are removed to some extent as it has increased the subjective as well as objective quality of the images. The algorithm effectively reduces the visibility of blocking artifacts along with the preservation of edges. It also increases the PSNR value of the image. As shown, the blocking artifacts are not removed totally. It is because of the fact that the information lost in the quantization step is irrecoverable. The algorithm only deals with pixel values (spatial domain) and the algorithm only tries to manipulate the pixel values on the basis of some criteria.

The extent of blocking artifacts can also be reduced by manipulating the DCT coefficients, as quantization is applied on the DCT coefficients. But frequency domain is having higher complexity and consumption of time. There is always a tradeoff between time (complexity) and efficiency (quality). Spatial domain is chosen where time (complexity) is the main concern, and on the other hand frequency domain is preferred where efficiency is given more value. The extent of reduction of blocking artifacts can be increased by recovering the information loss by using some sort of prediction algorithm. It can be done by some learning technique (artificial intelligence) or fuzzy logic. Further steps that can be taken in future are –

- The range of bit- rates can be extended in future.
- The postprocessor can be made generic postprocessor.
- It is possible to extend the proposed algorithm to video coding.
- Our proposed technique is restricted only to gray scale images, this can be extended to color images.
- Extension towards real-time compression by designing faster heuristic for estimating the interlayer dependencies.
- Design of a metric for measuring the blocking artifacts.

## REFERENCES

[1]    AHUMADA A. J.(1995) "Smoothing DCT Compression Artifacts" SID Digest, Vol. 25, pp. 708-711, 1995.

[2]    ARIA Nosratinia (2002) Department of Electrical Engineering, University of Texas at Dallas, "Enhancement of JPEG-Compressed Images by Re-application of JPEG" ©2002 Kluwer Academic Publishers. Printed in the Netherlands

[3]    AVERBUCH  Amir Z., SCHCLAR Alon and DONOHO David L. (2005) "Deblocking of Block-Transform Compressed Images Using Weighted Sums of Symmetrically Aligned Pixels" IEEE Transactions on Image Processing, Vol.14, No.2, February 2005.

[4]   CHANG Jyh-Yenng   and LU Shih-Mao ,"Image Blocking Artifact Suppression by the Modified Fuzzy Rule Based Filter" National Chiao Tung university, Hsinchu, Taiwan, ROC

[5]   GARG Ankush and RASIWASIA Nikhel ,Term Paper"Reducing Blocking Artifacts in Compressed Images" under Dr.Sumana Gupta,IIT Kanpur, pp. 6-10,15-16.

[6]   GONZALEZ Rafeal C. , WOODS Richard E , "Digital Image Processing", Second Edition, Pearson Publication.

[7]   GOTHUNDARUMUN A., WHITUKER RT and GREGOR J. (2001) " Total Variation for the Removal of Blocking Effects in DCT Based Encoding", 0-7803-6725- 1/01/$10.00 02001 IEEE

[8]   HYUK Choi and TAEJEONG Kim(2000) "Blocking-Artifact Reduction in Block-Coded Images Using Wavelet-Based Subband Decomposition", IEEE Transactions on Circuits and Systems for video technology, Vol. 10, No 5,August 2000.

[9]   ISMAEIL Ismaeil R. and WARD Rabab K. (2003) "Removal of DCT blocking Artifacts Using DC and AC Filtering" 0-7803-797tL0/03/$l7.00 © 2003 IEEE, pp.229-232.

[10]  KIRYUNG Lee, DONG Sik Kimand TAEJEONG Kim (2005)  "Regression-Based Prediction for Blocking Artifact Reduction in JPEG-Compressed Images"  IEEE Transactions on Image Processing, Vol.14, No.1, January 2005,pp-36-48.

[11]  KIZEWSKI Lukasz (2004) "Image Deblocking Using Local Segmentation" Paper for Bachelor of Software Engineering with Honours (2770) under  Dr. Peter Tischer, Monash  University ,November, 2004, kizewski@csse.monash.edu.au

[12]  KWON Kee-Koo , LEE Suk-Hwan ,LEE  Jong- Won, BAN Seong- Won , PARK Kyung-Nam , and LEE Kuhn-I1 (2001) "Blocking Artifacts Reduction Algorithm in Block Boundary area using Neural Network",0-7803-7090-2/0 1 /$ 10.00 © 2001 IEEE

[13]  LIEW Alan W.-C.  and YAN Hong (2004)  "Blocking Artifacts Suppression in Block-Coded Images Using Overcomplete Wavelet Representation", IEEE Transactions on Circuits and Systems for video technology, Vol 14,No 4,April 2004

[14]  LUO Ying and WARD Rabab K.   "Removing the Blocking Artifacts of Block-Based DCT Compressed Images"

[15]  NALLAPERUMAL Krishnan, RANJANI Jennifer J., CHRISTOPHER Seldev (2006) ,  "Removal of blocking artifacts in JPEG compressed images using Dual Tree Complex Wavelet Filters for Multimedia on the Web", 1-4244-0340-5/06/$20©2006IEEE

[16]  NIE Yao, KONG Hao-Song ,VETRO Anthony  and BARNER Kenneth (2005) "Fast Adaptive Fuzzy Post-Filtering for Coding Artifacts Removal in Interlaced Video" Mitsubishi Electric Research Laboratories, December 2005.

[17]  PARK Kyung-Nam , KWON Kee-Koo , BAN Seong- Won  and LEE Kuhn-I1  (2001) "Blocking Artifacts Reduction in Block-Coded Images Using Self-similarity",0-7803-7090-2/01/$10.00 0 2001 IEEE,pp.1667-1670.

[18]  ROBERTSON Mark A. and STEVONSON Robert L. "DCT Quantization Noise in Compressed Images"

[19]  TUAN Q., Pham Lucas J., VAN Vliet "Blocking artifacts removal by a hybrid filter method"

[20]  TRIANTAFYLLIDIS G.A., SAMPSON D., TZOVARAS D. and STRINTZIS  M.G. , "Blockiness Reduction in JPEG Coded Images",0-7803-7503-3/02/$17.00©2002 IEEE,pp.1325-1328.

[21] TRIANTAFYLLIDIS G.A., TZOVARAS D. and STRINTZIS M.G. (2001) "Detection of Blocking Artifacts of Compressed Still Images" 0-7695-1183-x/01 $10.00©2001 IEEE,pp.607-610.

[22] TRIANTAFYLLIDIS George A., TZOVARAS Dimitrios , SAMPSON Demetrios, Strintzis Michael G. (2002) "Combined Frequency and Spatial Domain Algorithm for the Removal of Blocking Artifacts" EURASIP Journal on Applied Signal Processing 2002:6, 601–612

[23] WANG Zhou, BOVIK Alan C., and EVANS Brian L. , " Blind Measurement of Blocking Artifacts in Images"

[24] WANG Ci,, ZHANG Wen-Jun, and FANG Xiang-Zhong (2004), "Adaptive Reduction of Blocking Artifacts in DCT Domain for Highly Compressed Images", 0098 3063/04/$20.00 © 2004 IEEE

[25] WANG Ci, XUE Ping , Weisi LIN, ZHANG Wenjun , and YU Songyu  (2005) "Fast Edge-Preserved Postprocessing for Compressed Images" IEEE Transactions on Circuits and Systems for Video Technology, Vol. 16, No. 9, September 2006,pp. 1142-1147.

[26] WENFENG  Gao, COSKUN Mermer, and YONGMIN Kim(2002) "A De-Blocking Algorithm and a Blockiness Metric for Highly Compressed Images" IEEE Transactions on Circuits and Systems for Video Technology, Vol. 12, No. 12, December 2002,pp.1150-1159.

[27] YANG Fu-zheng, WAN Shuai, CHANG Yi-lin, LUO Zhong (2006) "A no-reference blocking artifact metric for B-DCT video" Journal of Zhejiang University Received Nov. 30, 2005; revision accepted Feb. 17, 2006,ISSN 1009-3095 (Print); ISSN 1862-1775 (Online).pp.95-100.

[28] YANG Yongyi, GALATSANAS Nikolas P. and KATSAGGELAS Aggelas K., (1995) "Projection-Based Spatially Adaptive Reconstruction of Block Transform Compressed Images",IEEE Transaction on Image Processing, Vol4 as 7th July 1995,pp.421-431.

**Author**

Sukhpal Singh obtained his B.Tech. (Computer Science and Engineering) Degree from G.N.D.E.C. Ludhiana (Punjab) in 2010. He joined the Department of Computer Sci. & Eng. at North West Institute of Engineering and Technology, Moga (Punjab) in 2010. Presently he is pursuing M.E. (Software Engineering) degree from Thapar University,   Patiala. His research interests include Image Compression, Software Engineering, Cloud Computing, Operating System and Database.

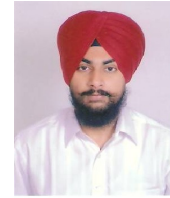